\documentclass[twocolumn,prb,superscriptaddress,floatfix,preprintnumbers,amssymb,amsmath]{revtex4-2}
\usepackage{graphicx}
\usepackage{dcolumn}
\usepackage{bm}
\usepackage[latin1]{inputenc}
\usepackage[mathscr]{eucal}
\usepackage{epsfig}

\begin{document}
\title{Down-conversion of quantum fluctuations of photonic heat current in a circuit}

\author{Bayan Karimi} 
\affiliation{Pico group, QTF Centre of Excellence, Department of Applied Physics, Aalto University School of Science, P.O. Box 13500, 00076 Aalto, Finland}
\email{bayan.karimi@aalto.fi, jukka.pekola@aalto.fi} 
\author{Jukka P. Pekola}
\affiliation{Pico group, QTF Centre of Excellence, Department of Applied Physics, Aalto University School of Science, P.O. Box 13500, 00076 Aalto, Finland}
\affiliation{Moscow Institute of Physics and Technology, 141700 Dolgoprudny, Russia}

\begin{abstract}
We discuss the non-zero frequency noise of heat current with the explicit example of energy carried by thermal photons in a circuit. Instead of the standard circuit modelling that gives a convenient way of predicting time-averaged heat current, we describe a setup composed of two resistors forming the heat baths by collections of bosonic oscillators. In terms of average heat transport this model leads to identical results with the conventional one, but besides this, it yields a convenient way of dealing with noise as well. The non-zero frequency heat current noise does not vanish in equilibrium even at zero temperature, the result that is known for, e.g., electron tunneling. We present a modulation method that can convert the difficult-to-measure high frequency quantum noise down to zero frequency.  
\end{abstract}

\date{\today}

\maketitle
\section{Introduction}
\begin{figure}
	\centering
	\includegraphics [width=\columnwidth] {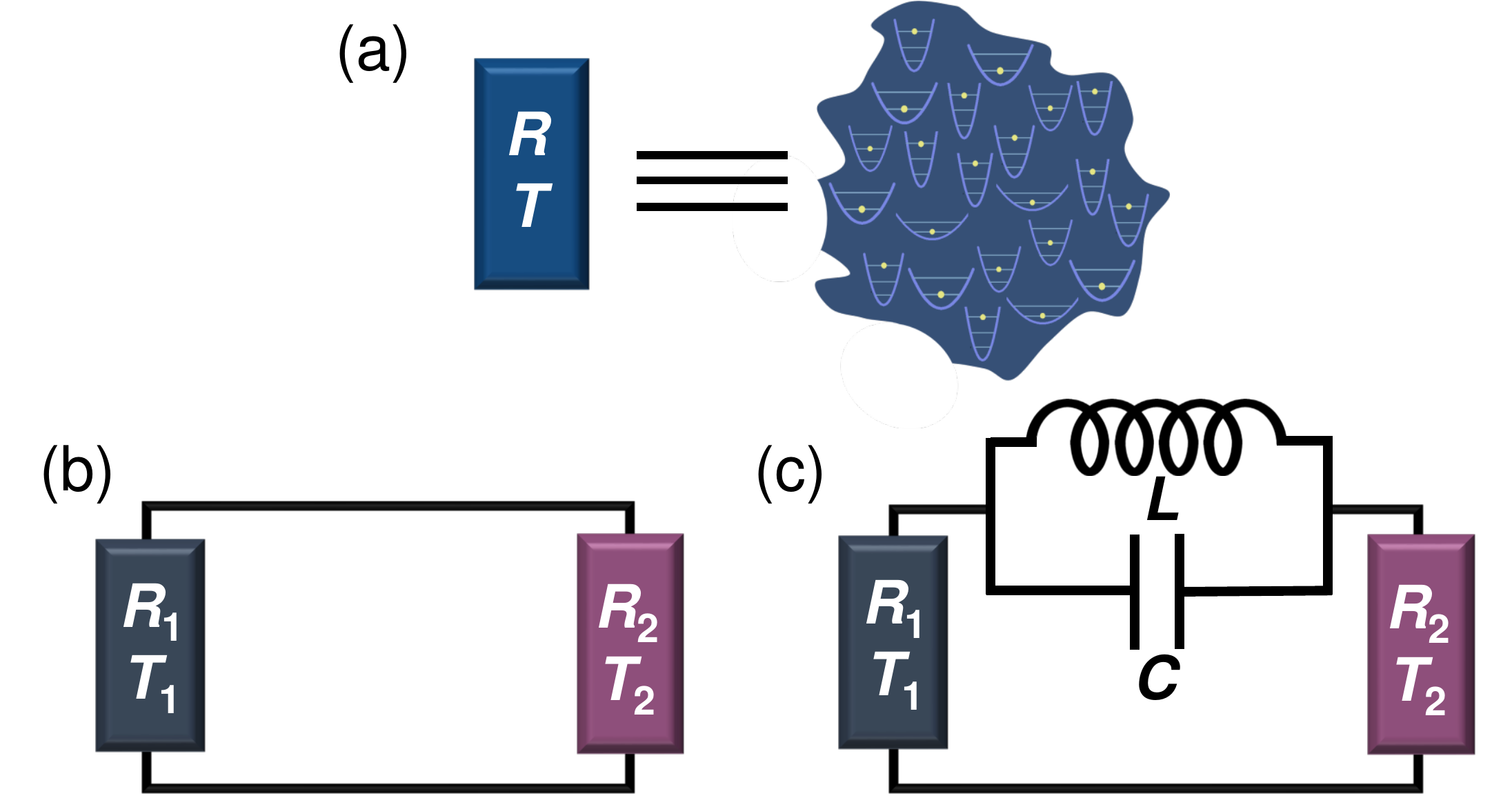}
	\caption{Elements for radiative heat transport in a circuit. (a) We model the resistors by a collection of bosonic oscillators. (b) The basic setup of two resistors $R_1$ and $R_2$ at temperatures $T_1$ and $T_2$, respectively. (c) Same as (b) but with a parallel $LC$-circuit with inductance $L$ and capacitance $C$ as frequency-dependent non-dissipative elements between the resistors, representing e.g. a classical superconducting quantum interference device (SQUID)~\cite{Tinkham}.
		\label{Fig1}}
\end{figure}

The quantum noise of heat current, i.e. noise at non-zero frequencies is an intriguing topic~\cite{Lifshitz,pekola1,Zhan,Sergi,AverinQT60,JB1,Crepieux}. It has been demonstrated theoretically for several processes, including heat transport by electrons, phonons and between the two, that this noise does not vanish even at zero temperature, thus jeopardizing the validity of fluctuation-dissipation theorem. Energy can be transported by various mechanisms. In condensed matter systems the most common carriers of energy in form of heat are electrons and phonons. Outside this domain radiation by photons in different frequency regimes provides an important channel of energy exchange and thermalization. But even within solid state systems,  radiation plays an important role, which is the topic of the current manuscript. In electrical circuits this mechanism corresponds to heat transmitted by thermal noise, which can also be interpreted as emission and absorption of thermal microwave photons~\cite{Schmidt,Meschke,Timofeev,Partanen}. In this context a straightforward way to treat the problem of average heat current in linear circuits is to employ a circuit theory where quantum noise emitted by dissipative elements induces Joule heating elsewhere in the said circuit~\cite{Schmidt,Pascal,GThomas}. To study noise of this heat current, various methods have been employed~\cite{Krive,GolubevPekola,Ciliberto}, including full-counting techniques with treatment of circuits by Keldysh Green functions but focusing on zero frequency noise. Here we address the archetypal Johnson-Nyquist setup of two resistors coupled to each other~\cite{Johnson1928,Nyquist1928} either directly or via a reactive element. We build up the resistors of bosonic oscillators~\cite{FV,Weiss,JB2,Brecht,Aurell}, as shown in Fig.~\ref{Fig1}(a), this way obtaining a microscopic Hamiltonian amenable for investigating noise at arbitrary frequencies, specifically finding the noise spectrum of heat in these configurations.
Observing the quantum heat current noise is an experimental challenge. To overcome it, we propose to shift the high frequency noise to low or zero frequency by modulating the coupling between the two heat baths periodically. This mixing principle, familiar for electrical measurements~\cite{Likharev,Koch,Lhotel}, has been proposed by Averin for fermionic heat noise~\cite{AverinQT60}. As pointed out by several authors~\cite{Gardiner}, it is important to specify what quantity to measure in order to make precise theoretical predictions. In this respect heat current noise after the down-conversion boils down to a measurement of time-dependent temperature of the mesoscopic heat bath (resistor).
Concretely we propose variation of a reactive element between the resistor baths to achieve this goal. 

\section{Modeling the system of two resistors}
The basic systems to be described in this work are shown in Fig.~\ref{Fig1}. We consider resistor $R_2$ with phase operator $\hat{\varphi}_2$ across it and subject to current operator $\hat{i}_2$ injection, due to the current produced by resistor $R_1$ arising from its thermal noise at temperature $T_1$. Its effective Hamiltonian, describing the system and its coupling to the bath $R_1$ is then 
\begin{equation} \label{e1}
\hat{H} = \hat{H}_2-\hat{i}_2\hat{\Phi}_2\equiv \hat{H}_2+\hat{\mathbb{V}},
\end{equation}
where $\hat{H}_2=\sum_k \hbar\omega_k \hat{c}_k^\dagger \hat{c}_k$ is the Hamiltonian of the bare resistor $R_2$ with oscillator energies $\hbar\omega_k$ and ladder operators $\hat{c}_k,\hat{c}_k^\dagger$, and $\hat{\Phi}_2=\frac{\hbar}{e}\hat{\varphi}_2$. 
To find the properties of the circuits we write the charge operator of the oscillators forming resistor $R_1$ and phase operator of the oscillators of $R_2$ as a linear combination with coefficients $\mu_j^{(1)}$ and $\lambda_k^{(2)}$ in the interaction picture as
\begin{eqnarray}\label{qphi1}
	\hat{q}_1(t)=\sum_j\imath\mu_j^{(1)}(\hat{b}_je^{-\imath\omega_jt}-\hat{b}_j^\dagger e^{\imath\omega_jt})
\end{eqnarray}	
\begin{eqnarray}\label{qphi2}	
	\hat{\varphi}_2(t)=\sum_k\lambda_k^{(2)}(\hat{c}_ke^{-\imath\omega_kt}+\hat{c}_k^\dagger e^{\imath\omega_kt}).
\end{eqnarray}
Here the ladder operators for resistor $R_1$ are $\hat{b}_j,\hat{b}_j^\dagger$, and the superscript (i) refers to the resistor $R_i$.

Let us consider the basic setup~\cite{Johnson1928,Nyquist1928} as shown in Fig.~\ref{Fig1}(b). We aim to calculate the photonic heat transport, its mean value and non-zero frequency noise, based on the model here. 
The current operator $\hat i_2$ through $R_2$ then reads $\hat{i}_2=\frac{R_1}{R_1+R_2}d\hat{q}_1/dt$, i.e., 
\begin{eqnarray}\label{I2}
	\hat{i}_2(t)=\frac{R_1}{R_1+R_2}\sum_j\mu_j^{(1)}\omega_j(\hat{b}_j^\dagger e^{\imath\omega_jt}+\hat{b}_j e^{-\imath\omega_jt}).
\end{eqnarray}
As a result, the coupling Hamiltonian $\hat{\mathbb{V}}$ is given by
\begin{eqnarray}\label{V}
	&&\hat{\mathbb{V}}(t)=-\frac{R_1}{R_1+R_2}\times\\&&\sum_{j,k}\lambda_k^{(2)}\mu_j^{(1)}\omega_j(\hat{b}_j^\dagger e^{\imath\omega_jt}+\hat{b}_j e^{-\imath\omega_jt})(\hat{c}_k^\dagger e^{\imath\omega_kt}+\hat{c}_k e^{-\imath\omega_kt}).\nonumber
\end{eqnarray}
This is in the form what one obtains by scattering theory imposing energy conservation~\cite{Fazio,pekola1}. In order to establish consistency between the oscillator bath model and the actual circuit, we request the current $\hat{i}=d\hat{q}_1/dt$ noise of resistor $R_1$, $S_i(\omega)=\int_{-\infty}^{\infty} dt e^{\imath\omega t}\langle \hat{i}(t)\hat{i}(0)\rangle$, to be equal to the quantum current noise of that resistor, i.e. $S_i(\omega)=\frac{2}{R_1}\frac{\hbar\omega}{1-\exp(-\beta_1\hbar \omega)}$, where $\beta_1=1/(k_{\rm B}T_1)$ is the inverse temperature of $R_1$.  Similarly we set the voltage $\hat{v}=d\hat{\Phi}_2/dt$ noise of resistor $R_2$, $S_v(\omega)=\int_{-\infty}^{\infty} dt e^{\imath\omega t}\langle \hat{v}(t)\hat{v}(0)\rangle$, equal to $S_v(\omega)=2R_2\frac{\hbar\omega}{1-\exp(-\beta_2\hbar \omega)}$ with similar notations. These conditions lead to the expressions
\begin{eqnarray}\label{mujlak}
	\mu_j^{(1)}=(\frac{\hbar}{\pi\nu_1(\omega_j)\omega_jR_1})^{1/2}, \,\,\, 	\lambda_k^{(2)}=(\frac{\hbar R_2}{\pi\nu_2(\omega_k)\omega_k})^{1/2},
\end{eqnarray}
where $\nu_i(\omega)$ is the oscillator density of states in $R_i$ with $i=1,2$. The operator of heat current to $R_2$, 
$\dot{\hat{H}}_2=\frac{\imath}{\hbar}[\hat{H},\hat{H}_2]=\frac{\imath}{\hbar}[\hat{\mathbb{V}},\hat{H}_2]$ reads then
\begin{eqnarray}\label{heatcurrent}
	\dot{\hat{H}}_2=\imath\frac{\hbar\sqrt{r_0}}{2\pi}&&\sum_{j,k}\frac{\sqrt{\omega_j\omega_k}}{\sqrt{\nu_1(\omega_j)\nu_2(\omega_k)}}\times\\&&(\hat{b}_j^\dagger e^{\imath\omega_jt}+\hat{b}_j e^{-\imath\omega_jt})(\hat{c}_k^\dagger e^{\imath\omega_kt}-\hat{c}_k e^{-\imath\omega_kt})\nonumber,
\end{eqnarray}
with $r_0=4R_1R_2/(R_1+R_2)^2$. 

\subsection{Quantum of thermal conductance}
According to the Kubo formula, the expectation value of heat current $\dot{Q}\equiv\langle \dot{\hat{H}}_2\rangle$ to $R_2$ under stationary conditions is
\begin{eqnarray}\label{heatcurrent1}
\dot{Q}=-\frac{\imath}{\hbar}\int_{-\infty}^{0}dt'\langle [\dot{\hat{H}}_2(0),\hat{\mathbb{V}}(t')]\rangle_0,
\end{eqnarray}
where $\langle.\rangle_0$ denotes the expectation value of a quantity for the non-interacting resistor at a given temperature. Since $\langle\hat{b}_j^\dagger\hat{b}_j\rangle_0=n_1(\omega_j)$ with $n_1(\omega)=1/(e^{\beta_1\hbar\omega}-1)$ the Bose-Einstein distribution of resistor $R_1$, and similarly $\langle\hat{c}_k^\dagger\hat{c}_k\rangle_0=n_2(\omega_k)$ for resistor $R_2$, the heat current is given by 
\begin{eqnarray}\label{heat}
\dot{Q}=r_0\frac{\pi k_{\rm B}^2}{12\hbar}(T_1^2-T_2^2),
\end{eqnarray}
which is equal to the expression obtained by standard circuit theory~\cite{Schmidt,Pascal,GThomas} thus providing a sanity check of our model. Equation \eqref{heat} yields heat conductance $G_{\rm th}=d\dot{Q}/dT_1|_T$ in form $G_{\rm th} = r_0 G_{\rm Q}$, where $G_{\rm Q} = \pi k_{\rm B}^2T/(6\hbar)$ is the thermal conductance quantum at temperature $T$~\cite{Pendry}. 

\subsection{Equilibrium quantum heat current noise}
The focus in this paper is the equilibrium noise~\cite{Callen} of heat current~\cite{Ciliberto,GolubevPekola}. With the same operators, this noise at (angular) frequency $\omega$, $S_{\dot{Q}}(\omega)=\int_{-\infty}^{\infty}e^{\imath \omega t}\langle\dot{\hat{H}}_2(t)\dot{\hat{H}}_2(0)\rangle$, is given by
\begin{eqnarray}\label{Somega6}
	S_{\dot{Q}}(\omega)=\frac{\hbar^2r_0}{2\pi}\int_{-\infty}^{\infty}d\Omega~\Omega(\Omega+\omega)n_1(\Omega)[1+n_2(\Omega+\omega)].\nonumber\\ 
\end{eqnarray}
The symmetrized heat current noise $S_{\dot{Q}}^{\rm symm}(\omega)\equiv (S_{\dot{Q}}(\omega)+S_{\dot{Q}}(-\omega))/2$ at equal temperatures $T_1=T_2\equiv T$ for two resistors reads
\begin{eqnarray}\label{Same}
	S_{\dot{Q}}^{\rm symm}(\omega)=\frac{r_0}{24\pi\hbar}[(2\pi k_{\rm B}T)^2+\hbar^2\omega^2]\hbar\omega\coth(\frac{\hbar\omega}{2k_{\rm B}T}).\nonumber\\ 
\end{eqnarray} 
This result exhibits non-vanishing noise at $\omega\neq 0$ even at $T=0$, a result known for some other systems~\cite{pekola1,Zhan,Sergi,AverinQT60,JB1} but not for the present one previously. At $\omega=0$ it reproduces the fluctuation-dissipation theorem~\cite{Callen,Lifshitz}. We note that identical results here and in what follows for the physical quantities $\dot{Q}$ and $S_{\dot{Q}}$ can be obtained with a similar strategy by replacing $\hat{i}_2\hat{\Phi}_2$ in Eq.~\eqref{e1} with $\hat{v}_2\hat{q}_2$, where $\hat{v}_2$ is the operator of voltage induced by $R_1$ on $R_2$, and $\hat{q}_2$ the charge operator for oscillators forming $R_2$.
  \begin{figure}
	\centering
	\includegraphics [width=\columnwidth] {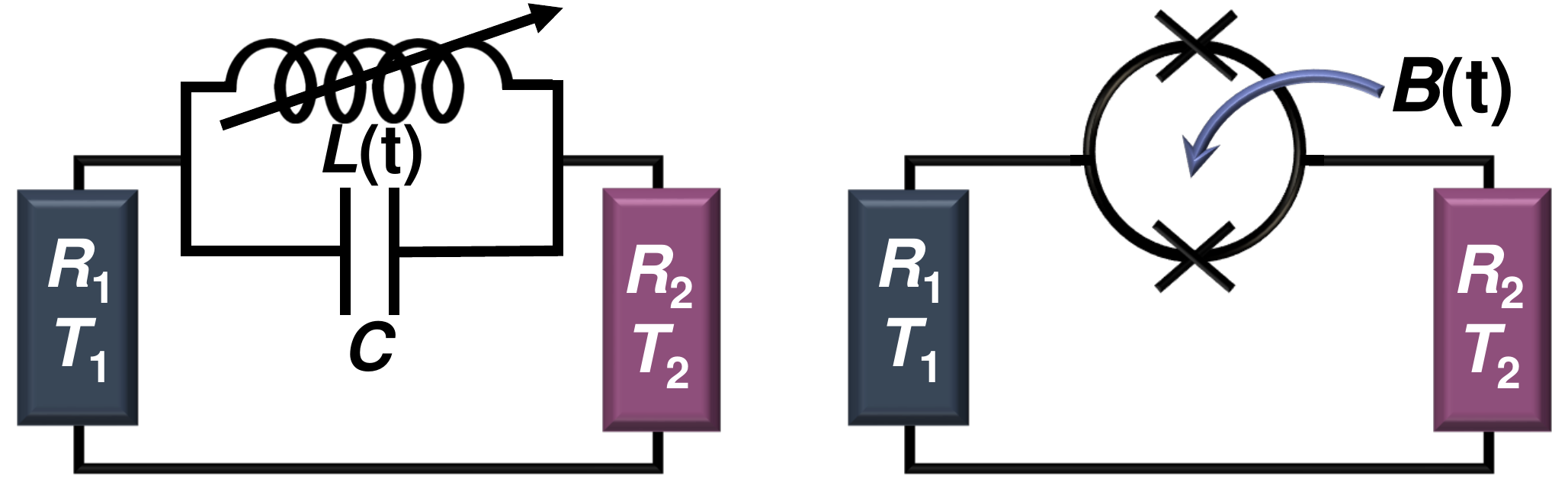}
	\caption{Illustration of the principle and a potential practical realization of quantum heat noise down-conversion. On the principal diagram on the left, the resistors are connected by a parallel $LC$-circuit, with variable inductor.This configuration can be realized as a SQUID, which is modulated by magnetic field $B(t)$. 
		\label{Fig2}}
\end{figure}
 
\section{Adding a reactive coupling element}
Next we investigate the influence of frequency dependent dissipationless impedance in the circuit. This element can then filter the heat current and it will be used as one of the examples in illustrating the noise down-conversion toward the end of the paper. Taking the circuit in Fig.~\ref{Fig1}(c) with parallel inductance $L$ and capacitance $C$, we find the differential equation for the current $\hat i_{\rm L}$ through the inductor given with the help of $\hat i=d\hat q_1/dt$ of the noise source with $\hat q_1$ from Eq.~\eqref{qphi1}, as $\frac{d^2i_{\rm L}}{dt^2}+\frac{1+\frac{C\dot{L}}{L}(R_1+R_2)}{C(R_1+R_2)}\frac{di_{\rm L}}{dt}+\frac{1}{LC}i_{\rm L}=\frac{1}{LC}\frac{R_1}{R_1+R_2}i(t)$. Here $\dot L$ is the rate of change of the inductance due to, e.g., variation of the magnetic flux through the superconducting quantum interference device (SQUID). For periodic sinusoidal variation of $L$ at driving frequency $\omega_0 /2\pi$, we have $|\dot L| \lesssim \omega_0 L$. Thus we may ignore the direct influence of $\dot L$ in this equation if $C\ll 1/(\omega_0(R_1+R_2))$. Since, as it will turn out later, the modulation frequency shall be of the order of $k_{\rm B}T/(2\pi \hbar)$ in order to obtain meaningful down-conversion, we have the condition $C \ll 2\pi \hbar/(k_{\rm B}T(R_1+R_2))$. This yields $C \ll 1\,$pF at $T=10\,$mK and for $R_1+R_2=100\,\Omega$, which is a forgiving bound since the typical junction capacitances are of the order of $1-10\,$fF. Then ignoring direct influence of $\dot L$, the solution of the circuit equation for $\hat i_2(t)$ is with the help of Eq.~\eqref{qphi2}
\begin{eqnarray}\label{i2-LC}
	i_2(t)=\frac{R_1}{R_1+R_2}&&\sum_{j}\mu_j^{(1)}\omega_j\times\\&&\{\frac{1-LC\omega_j^2}{(1-LC\omega_j^2)+\imath\frac{\omega_j L}{R_1+R_2}}\hat{b}_j^\dagger e^{\imath\omega_j t}+{\rm h.c.}\},\nonumber
\end{eqnarray}
which is exact for a stationary $LC$ circuit and approximately correct for sufficiently slowly varying $L$ as described above. Since the relevant angular frequencies $\omega_j$ are again of the order of $k_{\rm B}T/(2\pi \hbar)$, it turns out that this equation is well approximated by 
\begin{eqnarray}\label{i2_L}
	i_2(t)=\frac{R_1}{R_1+R_2}\sum_{j}\mu_j^{(1)}\omega_j\{\frac{1}{1+\imath\omega_j/\omega_{\rm L} }\hat{b}_j^\dagger e^{\imath\omega_j t}+{\rm h.c.}\},\nonumber\\
\end{eqnarray}
with the same condition for the capacitance $C$ as above. Here $\omega_{\rm L}=(R_1+R_2)/L$. This is the equation for a pure inductance $L$ between the two resistors, and exact for that case at arbitrary frequencies of modulation as well; thus at the end we only need to consider this simplified circuit both in the stationary and modulated case as long as the condition given for $C$ is satisfied. Using Eq.~\eqref{i2_L} and Eq.~\eqref{qphi2}, the coupling Hamiltonian is given by
\begin{eqnarray}\label{V2}
	\hat{\mathbb{V}}=-\frac{R_1}{R_1+R_2}&&\sum_{j,k}(\hat{c}_k^\dagger e^{\imath\omega_kt}+\hat{c}_k e^{-\imath\omega_kt})\times\\&&(\frac{\mu_j^{(1)}\omega_j}{1+\imath\omega_j/\omega_{\rm L}}\hat{b}_j^\dagger e^{\imath\omega_jt}+\frac{\mu_j^{(1)}\omega_j}{1-\imath\omega_j/\omega_{\rm L}}\hat{b}_j e^{-\imath\omega_jt}).\nonumber
\end{eqnarray}
The operator of heat current to $R_2$ for the circuit of Fig.~\ref{Fig1}(c) reads then
 \begin{eqnarray}\label{heatcurrent2}
 	\dot{\hat{H}}_2=\imath&&\frac{\hbar\sqrt{r_0}}{2\pi}\sum_{j,k}\frac{\sqrt{\omega_j\omega_k}}{\sqrt{\nu_1(\omega_j)\nu_2(\omega_k)}}(\hat{c}_k^\dagger e^{\imath\omega_kt}-\hat{c}_k e^{-\imath\omega_kt})\nonumber\\&&\times(\frac{1}{1+\imath\frac{\omega_j}{\omega_{\rm L}}}\hat{b}_j^\dagger e^{\imath\omega_jt}+\frac{1}{1-\imath\frac{\omega_j}{\omega_{\rm L}}}\hat{b}_j e^{-\imath\omega_jt}).
 \end{eqnarray}

\subsection{Average heat current}
The expectation value of the above operator which gives the heat current to $R_2$ is given by 
\begin{eqnarray}\label{Qdotinductor}
&&\dot{Q}=\int_{0}^{\infty}\frac{d\omega}{2\pi}\frac{4R_1R_2}{|R_1+R_2+\imath\omega L_0|^2}\hbar\omega[n_1(\omega)-n_2(\omega)],\nonumber\\
\end{eqnarray}
again obtained by Eq.~\eqref{heatcurrent1}. This is again the same result as that from the circuit theory, and it shows that the presence of non-vanishing inductance decreases the heat current, as expected. In order to solve the above integral analytically, we assume a small inductance such that $k_{\rm B}TL/\hbar \ll (R_1+R_2)$. Using the Taylor expansion for the integrand $4R_1R_2/|R_1+R_2+\imath\omega L|^2\simeq  r_0(1-(L/(R_1+R_2))^2\omega^2)$, the heat current between the two resistors via an inductor reads
\begin{eqnarray}\label{Qdotinductor2}
	\dot{Q}=r_0\frac{\pi k_{\rm B}^2}{12\hbar}(T_1^2-T_2^2)-r_0\frac{\pi^3 k_{\rm B}^4}{30\hbar^3}\frac{L^2}{(R_1+R_2)^2}(T_1^4-T_2^4).\nonumber\\
\end{eqnarray}
The first part is the heat current between two bare resistors of Eq.~\eqref{heat}. The thermal conductance then reads
\begin{equation}\label{thermalconduc2}
G_{\rm th}^{(\rm L)}=r_0G_{\rm Q}\big{\{}1-\frac{1}{5}(\frac{2\pi k_{\rm B}}{\hbar}\frac{L}{R_1+R_2})^2T^2\big{\}}.
\end{equation}
\subsection{Heat current noise}
With the same procedure as before, we obtain the symmetrized noise of heat current. In this circuit, we find that the lowest order correction to the result of Eq. \eqref{Same} is $-\delta S_{\dot{Q}}^{\rm symm}(\omega)$, where
\begin{eqnarray}\label{Sstatsymm}
	&&\delta S_{\dot{Q}}^{\rm symm}(\omega)=\frac{\hbar r_0}{4\pi}\frac{L^2}{(R_1+R_2)^2}\times\\&&\bigg{\{}\frac{1}{30}(\frac{2\pi k_{\rm B}T}{\hbar})^4+\frac{1}{12}(\frac{2\pi k_{\rm B}T}{\hbar})^2\omega^2+\frac{1}{20}\omega^4\bigg{\}}\hbar\omega\coth(\frac{\hbar\omega}{2k_{\rm B}T}).\nonumber 
\end{eqnarray}
Like the average thermal conductance, also the noise is suppressed by the inductive filter in between. 

\section{Experimental aspects}
We have shown that in both the configurations that we studied here, the noise of heat current at non-zero frequencies does not vanish even at zero temperature. A question arises whether these setups are realistic for experiment. The answer is positive: these configurations were proven to provide adequate description of the circuit in the experiments~\cite{Timofeev,Partanen} without including extra reactive elements modelling it. This is confirmed by the following estimates. The geometric inductance of a line of length 100 $\mu$m yields an inductance leading to impedance of $\sim 0.1\,\Omega$ at thermal frequencies at 100 mK. This impedance is well below a typical series resistance of $R_i=100$ $\Omega$. On the other hand, the parallel shunting capacitance for a similar circuit leads to an impedance of order 100 k$\Omega$ $\gg R_i$. Therefore, such parasitic impedances can be neglected in our analysis and in the basic experiments on micro-circuits at sub-kelvin temperatures. The same applies to the down-conversion measurements to be proposed below, since modulation frequencies are of the same order as temperature. Another question is how to observe the quantum heat current noise at high frequencies. This is a most challenging experiment; here we propose a mixing method  to shift the high frequency noise to low or zero frequency~\cite{AverinQT60}, where measuring such noise would be easier. 

\section{Down-conversion of heat current} 
According to the setup shown in Fig.~\ref{Fig2}, i.e. coupling two resistors via an inductor, the inductance is varied such that $L(t)=L_0[1+\eta\cos(\omega_0t+\phi)]$, where $\eta$ is the amplitude of the modulation and $\phi$ is the random phase of the drive with respect to system dynamics. In this case, we have the contribution from the ac drive as well. After averaging over $\phi$ the symmetrized heat current noise of the inductance modulation at frequency $\omega=0$, $S_{\dot{Q}}(0)$, at equal temperatures of the two resistors reads
\begin{equation}\label{Somega-Lmod9}
	S_{\dot{Q}}(0)=2k_{\rm B}T^2G_{\rm th}^{(\rm L)}+\frac{\eta^2}{2}\delta S_{\dot{Q}}^{\rm symm}(\omega_0),
\end{equation}
where $\delta S_{\dot{Q}}^{\rm symm}(\omega_0)$ is given in Eq.~\eqref{Sstatsymm} with $L=L_0$. Specifically for $T\rightarrow 0$, we find 
\begin{equation}\label{Somega-Lmod90}
	S_{\dot{Q}}(0)=\frac {\eta^2\hbar^2r_0}{160 \pi}\frac{L_0^2}{(R_1+R_2)^2}\omega_0^5.
\end{equation}

\section{Discussion}
The results obtained here for high frequency noise of heat current are in line with those derived for electron and phonon transport elsewhere. All these systems demonstrate non-vanishing noise at zero temperature, which is an intriguing property not easily accountable with the standard fluctuation-dissipation theorem. Not dwelling further on this last point, we focus finally on the experimental feasibility of observing these non-zero frequency fluctuations discussed above. The direct measurement of heat current noise is a challenging task even at low frequency, but much more so at high frequencies. Even the measurement principle for noise at these frequencies is not obvious: one needs to do it at GHz frequencies to make the frequency-dependent contribution dominant over the thermal one (see Eq. \eqref{Same}) even at a temperature of 10 mK achievable by standard techniques. The down-conversion of this noise, summarized by Eqs. \eqref{Somega-Lmod9} and \eqref{Somega-Lmod90}, can, however, be achieved in present-day experimental circuits operating at low temperatures. Varying the reactive impedance in between fixed resistors (Fig.~\ref{Fig2}) could be realized by a standard technique, by varying the magnetic flux through a SQUID that acts as a tunable inductor. Another option, not analyzed here, would be to vary the resistance of one of the two resistors. Since the electrical conductance of two-dimensional materials can be easily controlled by gate voltage due to Coulomb repulsion, the resistors could be formed either out of semiconducting two-dimensional electron gas~\cite{Rimberg} or out of graphene. Finally, the measurement of heat current noise can in practise be realized as in Ref.~\cite{BJ}, by detecting the variations of the effective temperature of a nanocalorimeter using a fast tunnel junction thermometer.   

We acknowledge valuable discussions with Dmitri Averin and useful inputs from Sergey Kubatkin. This work was supported by Academy of Finland grant 312057 and by the European Union's Horizon 2020 research and innovation programme under the European Research Council (ERC) programme (grant agreement 742559). We thank the Russian Science Foundation (Grant No. 20-62-46026) and Foundational Questions Institute Fund (FQXi) via Grant No. FQXi-IAF19-06 for supporting the work.

\end{document}